\documentclass [11pt]{article}
\title{Evolution and Earth's Entropy}
\author{Robert D. Klauber\\ Am. J. Phys. \textbf{77}(9), 773-774 (2009)\\rklauber(AT)netscape.net}

\date{June 4, 2009}

\usepackage{graphicx}
\usepackage{longtable}

\begin{document}
\maketitle

I greatly enjoyed the interesting insights in Styer's\cite{Styer:2008}$^{ 
}$ and Bunn's\cite{Bunn:2009} articles on entropy and evolution and would 
like to add a couple of points.

Styer showed the total entropy throughput rate for the Earth to be about 
$4\times 10^{14}\,\mbox{J/K}\,\mbox{s}$ and estimated a maximum rate of 
decrease in entropy from evolution of the order of $3\times 
10^2\,\mbox{J/K}\,\mbox{s}$. Bunn did a more robust analysis of evolutionary 
entropy change and determines it to be far less than 10$^{21}$ J/K over the 
history of the Earth. For a $4.5\times 10^9$ year old Earth, this is about 
10$^{4}$ J/K s.

Styer noted the evolutionary entropy decrease is many orders of magnitude 
below the Earth entropy throughput, and this holds even for Bunn's estimate. 
Both authors took, as their starting point, the principle that $\Delta S\ge 
0$ for an isolated system and identified that system to be the Earth plus 
those objects with which the Earth exchanges energy. They then showed that 
the total entropy of this system increases, even when the estimates for 
evolutionary decreases are included. However, neither discussed the Earth 
itself as a system (not thermally isolated) and the mechanism for the 
overall entropy decrease within that system, which underlies the growth of 
all life as we know it.

The decrease in entropy for the Earth system can be attributed to the 
incoming heat from the sun occurring at higher temperature on~the~planet's 
surface during the~day, whereas the same amount of heat (approximately, on 
average) leaving the planet during the subsequent night is at a lower 
temperature.~Hence, $\left| {\Delta Q/T_{in} } \right|<\left| {\Delta 
Q/T_{out} } \right|$ and Earth's entropy decreases.~A quick estimate, 
assuming 5\r{ }C average variation between night and day,~shows this to be 
on the order of 2{\%} of the Earth entropy throughput, i.e., greater than 
10$^{12}$ J/K s. Nonequilibrium energy exchanges solely on and within the 
Earth also produce entropy, effectively reducing the net decrease to 
something near the relatively miniscule levels Styer and Bunn 
suggested.\cite{For:1}\cite{Epstein:1999}

I do note that the total entropy decrease~should not correlate solely with 
the existence of more complex organisms than at earlier times, which Styer 
discussed, but also to a greater total number of organisms.~The human race, 
for example, has over 5 $\times $ 10$^{9}$ more individuals now than it had 
a couple of centuries ago, and all of these are far more organized than the 
base elements and compounds were before they formed those individuals' 
bodies.

Further, it seems that living creatures, particularly human ones, organize 
things around themselves into lower entropy configurations that last beyond 
the lives of those creatures (whose deaths lead to local~increases in 
entropy).~ Cathedrals erected from stone, and societies emerging from clans, 
are two examples, though in the latter case, I may, as Styer warned against, 
be~intermingling the metaphor with the definition.

None of this contradicts anything in the articles. The entropy decreases 
involved in evolution, by any estimate, remain many orders of magnitude less 
than the total entropy gain of the universe, the total Earth entropy 
throughput, and importantly, the net planetary entropy decrease from 
day/night temperature differences. This trumps the creationist argument, 
addressed by both authors, that claims science is somehow internally 
inconsistent with regard to evolution.


\begin{thebibliography}{3}
\bibitem{For:1} For a constant volume, closed system in thermal contact with its surroundings, $\Delta S=\sum\limits_{i\,} {\left( {\Delta Q/T} \right)_i } +\Delta S_{internal} $, where the last term is due to nonequilibrium processes and is always positive. For the Earth, the summation terms are, effectively, $\left| {\Delta Q} \right|/T_{in} -\left| {\Delta Q} \right|/T_{out} <0$, which can, and do, lead to $\Delta S<0$ within the system.
\end{thebibliography}
\end{document}